\documentclass[aps,prc,reprint,superscriptaddress,amsmath,amssymb,floatfix]{revtex4-2}

\usepackage{CJK}
\usepackage{graphicx}
\usepackage{picture}
\usepackage{dcolumn}
\usepackage{bm}
\usepackage{subcaption}
\usepackage{ragged2e}
\usepackage[bookmarks=true,colorlinks=true,
citecolor=blue,linkcolor=blue,anchorcolor=blue,filecolor=blue,urlcolor=blue]{hyperref}
\usepackage{pgfplots}
\usepackage{booktabs}
\usetikzlibrary{patterns}
\begin{document}
\begin{CJK*}{UTF8}{gbsn}
	
\preprint{APS/123-QED}

\title{Radiative decay branching ratio of the Hoyle state}

\author{Zifeng Luo (罗梓锋)}
\email{luozf@tamu.edu}
\affiliation{Department of Physics \& Astronomy, Texas A\&M University, College Station, Texas 77843, USA}
\affiliation{Cyclotron Institute, Texas A\&M University, College Station, Texas 77843, USA}
\author{M.~Barbui}
\affiliation{Cyclotron Institute, Texas A\&M University, College Station, Texas 77843, USA}
\author{J.~Bishop}
\thanks{Present address: University of Birmingham, Edgbaston, B15 2TT, UK}
\affiliation{Cyclotron Institute, Texas A\&M University, College Station, Texas 77843, USA}
\author{G.~Chubarian}
\affiliation{Cyclotron Institute, Texas A\&M University, College Station, Texas 77843, USA}
\author{V.Z. ~Goldberg}
\affiliation{Cyclotron Institute, Texas A\&M University, College Station, Texas 77843, USA}
\author{E.~Harris}
\affiliation{Department of Physics \& Astronomy, Texas A\&M University, College Station, Texas 77843, USA}
\affiliation{Cyclotron Institute, Texas A\&M University, College Station, Texas 77843, USA}
\author{E.~Koshchiy}
\affiliation{Cyclotron Institute, Texas A\&M University, College Station, Texas 77843, USA}
\author{C.E.~Parker}
\affiliation{Cyclotron Institute, Texas A\&M University, College Station, Texas 77843, USA}
\author{M.~Roosa}
\affiliation{Department of Physics \& Astronomy, Texas A\&M University, College Station, Texas 77843, USA}
\affiliation{Cyclotron Institute, Texas A\&M University, College Station, Texas 77843, USA}
\author{A. Saastamoinen}
\affiliation{Cyclotron Institute, Texas A\&M University, College Station, Texas 77843, USA}
\author{D.P.~Scriven}
\affiliation{Department of Physics \& Astronomy, Texas A\&M University, College Station, Texas 77843, USA}
\affiliation{Cyclotron Institute, Texas A\&M University, College Station, Texas 77843, USA}
\author{G.V. Rogachev}
\email[Corresponding author: ]{rogachev@tamu.edu}
\affiliation{Department of Physics \& Astronomy, Texas A\&M University, College Station, Texas 77843, USA}
\affiliation{Cyclotron Institute, Texas A\&M University, College Station, Texas 77843, USA}
\affiliation{Nuclear Solutions Institute, Texas A\&M University, College Station, Texas 77843, USA}

\begin{abstract}
	\noindent{\bf Background:}
    The triple-alpha process is a vital reaction in nuclear astrophysics, characterized by two consecutive reactions [$2\alpha\leftrightarrows{^{8}\rm{Be}}(\alpha,\gamma){^{12}\rm{C}}$] that drive carbon formation. The second reaction occurs through the Hoyle state, a 7.65 MeV excited state in ${^{12}\rm{C}}$ with $J^{\pi}=0^{+}$. The rate of the process depends on the radiative width, which can be determined by measuring the branching ratio for electromagnetic decay. Recent measurements by Kib\'edi \textit{et al.} conflicted with the adopted value and resulted in a significant increase of nearly 50\% in this branching ratio, directly affecting the triple-alpha reaction.
	\\
	\noindent{\bf Purpose:}
	This work aims to utilize charged-particle spectroscopy with magnetic selection as a means to accurately measure the total radiative branching ratio ($\Gamma_{\rm{rad}}/\Gamma$) of the Hoyle state in $^{12}{\rm C}$.
	\\
	\noindent{\bf Methods:}
	The Hoyle state in $^{12}{\rm C}$ was populated via $^{12}\rm{C}(\alpha, \alpha')^{12}\rm{C^{*}}$ inelastic scattering. The scattered $\alpha$ particles were detected using a $\Delta$E-E telescope, while the recoiled $^{12}{\rm C}$ ions were identified in a magnetic spectrometer.
	\\
	\noindent{\bf Results:}
	A radiative branching ratio value of $\Gamma_{\rm{rad}}/\Gamma\times10^{4}=4.0\pm0.3({\rm stat.})\pm0.16({\rm syst.})$ was obtained.
	\\
	\noindent{\bf Conclusions:}
	The radiative branching ratio for the Hoyle state obtained in this work is in agreement with the original adopted value. Our result suggests that the proton-$\gamma$-$\gamma$ spectroscopy result reported by Kib\'edi \textit{et al.} may be excluded.
\end{abstract}

\pacs{Valid PACS appear here}


\maketitle
\end{CJK*}

\section{Introduction}
The triple-alpha process is a crucial reaction in the field of nuclear astrophysics, consisting of two consecutive reactions, (a) $\alpha+\alpha\rightarrow{^{8}\rm{Be(g.s.)}}$ and (b) $^{8}\rm{Be} +\alpha\rightarrow\gamma+{^{12}\rm{C}}$, ultimately leading to the formation of carbon. The second reaction occurs via an excited $0^{+}$ state at an excitation energy of 7.65 MeV in $^{12}{\rm C}$, known as the Hoyle state \cite{hoyle1954nuclear}. This state predominantly decays by $\alpha$ emission, but a small branch of electromagnetic decay ultimately forms the $^{12}{\rm C}$(g.s.). The rate of the triple-alpha process is determined by the product of the $\alpha$ decay width ($\Gamma_{\alpha}$) and the radiative width ($\Gamma_{\rm{rad}}$) divided by their sum ($\Gamma_{\alpha}+\Gamma_{\rm{rad}}$). As shown in Eq. \ref{EQU1}, the experimental method of determining the value of $\Gamma_{\rm{rad}}$ involves measuring the branching ratio for electromagnetic decay ($\Gamma_{\rm{rad}}/\Gamma$) and utilizing the established partial width $\Gamma_{\pi}(\rm{E0})$ for electron-positron pair production \cite{Kibedi2020}. 
\begin{equation}\label{EQU1}
    \Gamma_{\rm{rad}}=\frac{\Gamma_{\rm{rad}}}{\Gamma}\times\frac{\Gamma}{\Gamma_{\pi}(\rm{E0})}\times\Gamma_{\pi}(\rm{E0})
\end{equation}

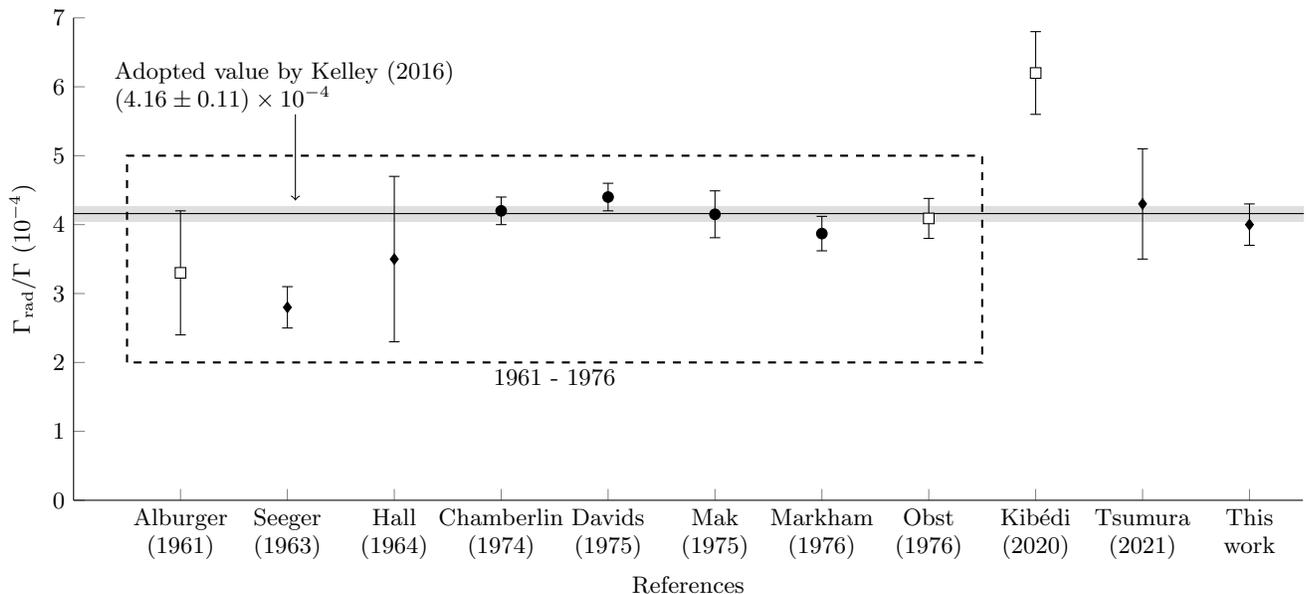
\begin{figure*}[htbp]
	\centering
	\captionsetup{justification=RaggedRight}
	 \usepgfplotslibrary{fillbetween}
	\begin{tikzpicture}
		\begin{axis}[
                height=8cm,
                width=\textwidth,
			xtick pos=left,
			ytick pos=left,
			axis y line*=left,
			axis x line*=left,
			ylabel near ticks,
			xlabel near ticks,
			xlabel=References,
			ylabel=$\Gamma_{\rm{rad}}/\Gamma\ (10^{-4})$,
			xmin=0,xmax=11.5,
			ymin=0, ymax=7,
			ytick={0,1,2,3,4,5,6,7},
			xtick={1,2,3,4,5,6,7,8,9,10,11,12,13},
			xticklabels={Alburger\\(1961),Seeger\\(1963),Hall\\(1964),Chamberlin\\(1974),Davids\\(1975),Mak\\(1975),Markham\\(1976),Obst\\(1976),Kib\'edi\\(2020),Tsumura\\(2021),This\\work},
            xticklabel style={align=center},
			]
			
			\node[anchor=mid,align=right] at (axis cs:4.5,1.75){1961 - 1976};
            \draw [thick,dashed] (5,200) rectangle (85,500);
   
			\addplot[mark=none, black] coordinates {(0,4.16) (13,4.16)};
			\addplot[mark=none, draw opacity=0, name path=UpError] coordinates {(0,4.27) (13,4.27)};
			\addplot[mark=none, draw opacity=0, name path=LowError] coordinates {(0,4.05) (13,4.05)};
			\addplot[gray!25] fill between[of=UpError and LowError, , soft clip={domain=0.015:14}];
			
			\node[anchor=west,align=left] at (axis cs:0.3, 6){Adopted value by Kelley (2016)\\$(4.16\pm0.11)\times10^{-4}$};
			\addplot[->] coordinates {(2.075,5.6) (2.075,4.35)};
			
			\addplot[
			scatter/classes={a={white}, b={black}},
			scatter,
			only marks,
			mark=square*,
			mark options={fill=white},
			]
			plot [error bars/.cd, y dir = both, y explicit]
			table[meta=class, x=x, y=y, y error=ey]{
				x   y    ey    class
				1   3.3  0.9   b      
				8  	4.09 0.29  b      
				9   6.2  0.6   b      
			};
		
			\addplot[
			scatter/classes={a={white}, b={black}, c={blue}},
			scatter,
			only marks,
			mark=*,
			mark options={fill=black},
			]
			plot [error bars/.cd, y dir = both, y explicit]
			table[meta=class, x=x, y=y, y error=ey]{
				x   y    ey    class  
				4   4.2	 0.2   b
				5   4.4	 0.2   b
				6   4.15 0.34  b
				7   3.87 0.25  b
			};
            \addplot[
			scatter/classes={a={white}, b={black}, c={blue}},
			scatter,
			only marks,
			mark=diamond*,
			mark options={fill=black},
			]
			plot [error bars/.cd, y dir = both, y explicit]
			table[meta=class, x=x, y=y, y error=ey]{
				x   y    ey    class
				2   2.8  0.3   b
                3  	3.5  1.2   b
				10  4.3  0.8   b
                11  4.0  0.3   b
			};
		\end{axis}
	\end{tikzpicture}
	\caption{Summary of $\Gamma_{\rm{rad}}/\Gamma$ measurements from  Alburger (1961) \cite{Ref1961}, Seeger (1963) \cite{Ref1963-1}, Hall (1964) \cite{Ref1963-2}, Chamberlin (1974) \cite{Ref1974}, Davids (1975) \cite{Ref1975-1}, Mak (1975) \cite{Ref1975-2}, Markham (1976) \cite{Ref1976-1}, Obst (1976) \cite{Ref1976-2}, Kib\'edi (2020) \cite{Kibedi2020}, Tsumura (2021) \cite{Tsumura2021}, and this work. The black diamonds and circles represent measurements conducted using charged-particle spectroscopy with and without magnetic selection, respectively, while the white boxes represent measurements employing $\gamma$-particle coincidence methods. The value from Ref. \cite{Ref1963-1} has been excluded from the adopted value by Kelly (2016) \cite{KELLEY201771} as it is considered a statistical outlier.}
	\label{fig:results}
\end{figure*}

The radiative branching ratio ($\Gamma_{\rm{rad}}/\Gamma$), the first parameter on the right-hand side of Eq. \ref{EQU1}, has gathered significant attention over the years. Previous studies conducted during the 20th century \cite{Ref1961,Ref1963-1,Ref1963-2,Ref1974,Ref1975-1,Ref1975-2,Ref1976-1,Ref1976-2} contributed to an adopted value of $\Gamma_{\rm{rad}}/\Gamma=4.16(11)\times10^{-4}$ \cite{KELLEY201771}. However, a recent study by Kib\'edi (2020) \cite{Kibedi2020} has unveiled a substantial deviation exceeding 3$\sigma$ from the adopted value, introducing considerable uncertainties in the determination of the triple-alpha reaction rate. In contrast, the measurement conducted by Tsumura (2021) \cite{Tsumura2021} exhibits agreement with previous findings, although it still possesses a relatively large uncertainty. Figure \ref{fig:results} provides a comprehensive overview of these measurements. The observed discrepancies emphasize the importance of further investigating the radiative branching ratio of the Hoyle state, as it has a profound impact on various astrophysical model calculations, shaping our understanding of stellar nucleosynthesis and the formation of carbon in the universe. This paper introduces a new measurement of $\Gamma_{\rm{rad}}/\Gamma$ using the charged-particle spectroscopy method with magnetic selection, effectively resolving the ambiguity associated with the radiative decay branching ratio of the Hoyle state in $^{12}{\rm C}$.

\section{Experiment}\label{sec:experiment}
\begin{figure}[htbp] 
\centering 
\captionsetup{justification=RaggedRight}
\includegraphics[width=0.48\textwidth]{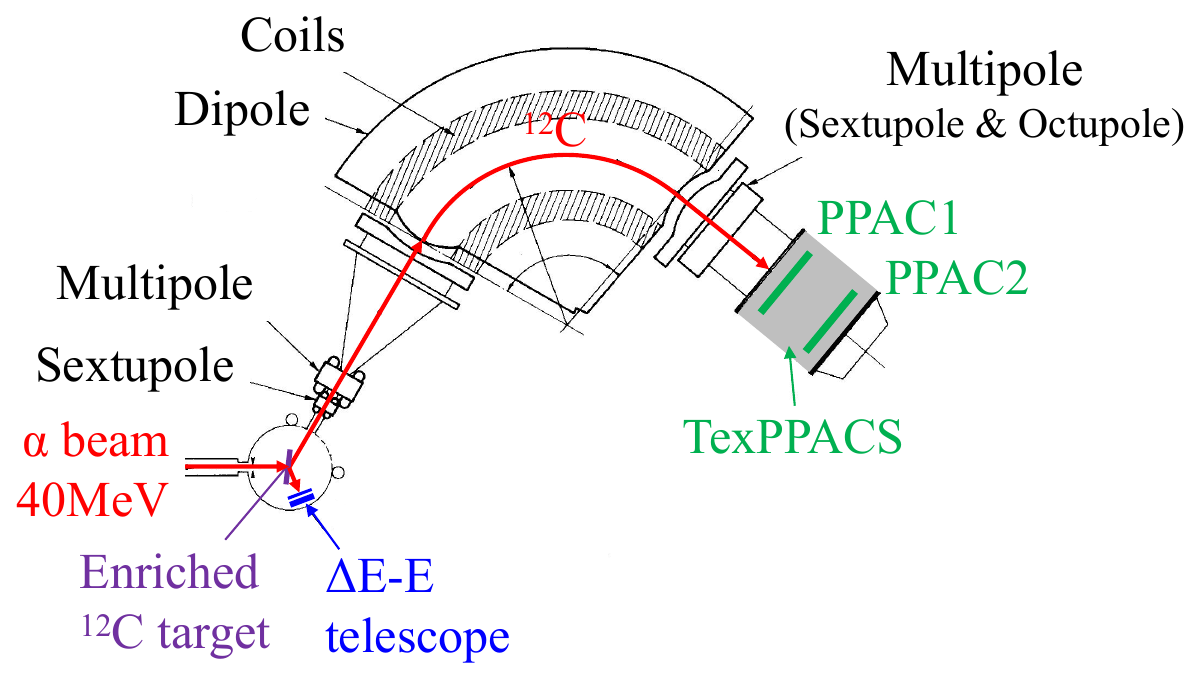}
\caption{Schematics of the experimental setup.}
\label{fig:setup}
\end{figure}

The experiment was conducted at the Cyclotron Institute at Texas A\&M University, using the K150 Cyclotron. Figure \ref{fig:setup} illustrates the experimental setup employed in this study. The excited states in $^{12}{\rm{C}}$ were populate by inelastic scattering of 40 MeV $\alpha$ particles on a highly enriched $^{12}{\rm{C}}$ target provided by Argonne National Laboratory. It contains less than 0.17\% (verified experimentally; see discussion in Sec. \ref{sec:pid}) of $^{13}{\rm{C}}$ and has a thickness of 159 $\mu$g/cm$^{2}$.

The scattered $\alpha$ particles were detected and identified using a $\Delta$E-E silicon telescope, consisting of a 32-$\rm\mu$m-thick silicon detector and a 500-$\mu$m-thick double-sided silicon strip detector (DSSD) manufactured by Micron Semiconductor Ltd \cite{Micron}. Both silicon detectors have an active area of 49.5 mm $\times$ 49.5 mm. The front and rear sides of the DSSD were divided into 16 vertical strips and 16 horizontal strips, respectively. Positioned at an angle of 81.15$^{\circ}$ relative to the beam axis and located 14.2 cm away from the target, the silicon telescope covered from 72.8$^{\circ}$ to 89.5$^{\circ}$ on the reaction plane. This setup facilitated the determination of the momentum vector of the $\alpha$ particles.

To detect the $^{12}{\rm{C}}$(g.s.) ions surviving from the electromagnetic decay of the $^{12}{\rm{C}}(0_{2}^{+})$ recoil, we employed the Multipole-Dipole-Multipole (MDM) spectrometer \cite{MDM}. Positioned at an angle of 35.3$^{\circ}$ in the laboratory frame, the MDM spectrometer covered 4$^{\circ}$ in both the vertical and horizontal directions, providing a large acceptance for the $^{12}{\rm C}$ recoil ions. The detection and identification of the recoil ions, filtered by the MDM spectrometer, were facilitated by implementing the Texas Parallel-Plate Avalanche Counter System (TexPPACS) at the end of MDM. The TexPPACS featured a 2.5 $\rm\mu$m Mylar entrance window, operated with 4 Torr pentane gas and spaced at a distance of 42 cm, allowing for the efficient detection of heavy ions with energies around 1 MeV/nucleon. The time between the DSSD and each of the two PPACs in the TexPPACS detector system was recorded. The effective timing resolution of the TexPPACS was determined to be 1.5 ns. 

A dual-trigger mode was implemented to accommodate the high counting rate. The trigger output from the DSSD's front strips shaper was divided into two channels. One channel was used for coincidences with the first PPAC detector, while the other channel was connected to a prescaler that generated one output signal for every 100 input signals. The coincidence and the scaler output triggered the data acquisition system independently. This dual trigger mode significantly reduced the counting rate for single events by a factor of 100 while ensuring the capture of all $\alpha$+$^{12}{\rm C}$ coincidence events.

The branching ratio can be calculated using the following expression:
\begin{equation}\label{ratio-1}
\frac{\Gamma_{\rm{rad}}}{\Gamma}=\frac{N_{\rm{Coinc}}}{100\times N_{\rm{Scaled}} \times F_{5^+} \times \epsilon},
\end{equation}
where $N_{\rm{Coinc}}$, $N_{\rm{Scaled}}$, $F_{5+}$, and $\epsilon$ represent the yield of coincidence events, scaled single events, the charge state fraction of $^{12}{\rm C}^{5+}$ (see discussion in Sec. \ref{sec:charge}), and the efficiency that accounts for various factors such as MDM-TexPPACS efficiency, target thickness, etc. respectively. 

\section{Analysis}

In this section, we present an analysis detailing the steps taken to determine the radiative decay branching ratio for the Hoyle state. We begin by discussing the charge state fraction distribution of $^{12}{\rm C}$, which serves as the starting point of our analysis.

\subsection{Charge state fraction distribution}\label{sec:charge}
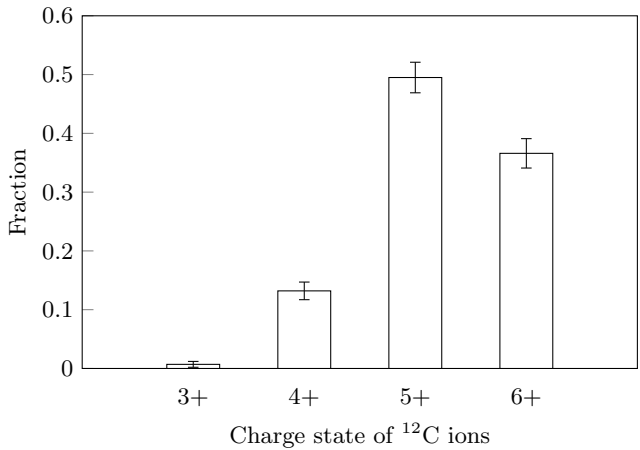
\begin{figure}[htbp]
	\centering
	\captionsetup{justification=RaggedRight}
	\begin{tikzpicture}
		\begin{axis}[
		width  = 0.5*\textwidth,
		height = 0.35*\textwidth,
		major x tick style = transparent,
		ybar=0.5*\pgflinewidth,
		bar width=20pt,
		ymajorgrids = false,
		xtick pos=left,
		ytick pos=left,
		ylabel near ticks,
		xlabel near ticks,
		xlabel=Charge state of $^{12}{\rm C}$ ions,
		ylabel=Fraction,
		xmin=0, xmax=5,
		ymin=0, ymax =0.6,
		ytick={0,0.1,0.2,0.3,0.4,0.5,0.6},
		xtick={1,2,3,4},
		xticklabels={3+, 4+, 5+, 6+},
		]
		\addplot[style={fill=white},error bars/.cd, y dir=both, y explicit]
			coordinates {
			(1, 0.007) += (0,0.005) -= (0,0.005)
			(2, 0.132) += (0,0.015) -= (0,0.015)
			(3, 0.495) += (0,0.026) -= (0,0.026)
			(4, 0.366) += (0,0.025) -= (0,0.025)
			};
		\end{axis}
	\end{tikzpicture}
	\caption{Charge state fraction distribution of $^{12}{\rm C}$ ions leaving the target.}
	\label{fig:chargeState}
\end{figure}

The $^{12}{\rm C}$ ions emitted from the target exhibit charge states ranging from 1$^+$ to 6$^+$. In order to ascertain the distribution of $^{12}{\rm C}$ charge states after the target, the magnetic rigidity of the MDM was adjusted individually for $^{12}{\rm C}$ in each charge state. We used the first excited state of $^{12}{\rm C}$, the 2$^+$ at 4.44 MeV, to record the coincidence between the respective $\alpha$ particles in the DSSD and the $^{12}{\rm C}$ in the TexPPACS. The resulting distribution is shown in Fig. \ref{fig:chargeState}. It can be observed that $^{12}{\rm C}^{5^+}$ holds the highest fraction, with $F_{5+} = 0.495\pm0.026$. Therefore, we selected the magnetic rigidity of the MDM spectrometer for the $^{12}{\rm C}^{5+}$ charge state. This also eliminates most of the $\alpha$ particles originating from the 3$\alpha$ decay of the Hoyle state.

\subsection{Efficiency}
The total efficiency of the MDM-TexPPACS system was estimated using Monte Carlo simulation with Geant4 \cite{Geant4} + RAYTRACE \cite{RAYTRACE}. This comprehensive simulation considered various experimental factors, such as beam emittance, target thickness, transmission efficiency of the MDM spectrometer, and geometrical efficiency for coincidence selection. The accuracy of the simulation was validated by comparing the simulated efficiencies and the experimentally-determined efficiencies of the setup for measuring the radiative branching ratio of $^{12}{\rm C}(2^{+}_{1})$, which is known to be 100\%. 

To investigate the impact of beam emittance, which is approximately 24 mm-rad for the K150 cyclotron, we conducted a series of simulations based on different emittance distributions. The average of the simulated results was considered as the efficiency value, while the largest difference observed among them was taken as a source of systematic uncertainty.

The resulting efficiency value for the setup targeting $^{12}{\rm C}(0^{+}_{2})$ was determined to be $\epsilon=0.95\pm0.03 ({\rm syst.})$.

\subsection{Particle identification\label{sec:pid}}
The magnetic rigidity of the MDM spectrometer was optimized to select the heavy recoils of interest: the $^{12}{\rm C}$ ions resuting from the population of the Hoyle state in $\alpha$ particle inelastic scattering on $^{12}{\rm C}$. Other recoil ions with similar magnetic rigidity produced in the interaction of the $\alpha$ particle beam with the target ($^{13}{\rm C}$, $^{16}{\rm O}$, and $\alpha$ particles) can pass through the spectrometer as well. The time of flight (ToF), defined as the difference ($T_{1}-T_{\rm Si}$) between the ToF of the ions from the target to the silicon detector, $T_{\rm Si}$, and the ToF of the ions from the target to the first PPAC, $T_{1}$, offers good particle identification.

\begin{figure}[htbp] 
\centering
\captionsetup{justification=RaggedRight}
\includegraphics[width=0.48\textwidth]{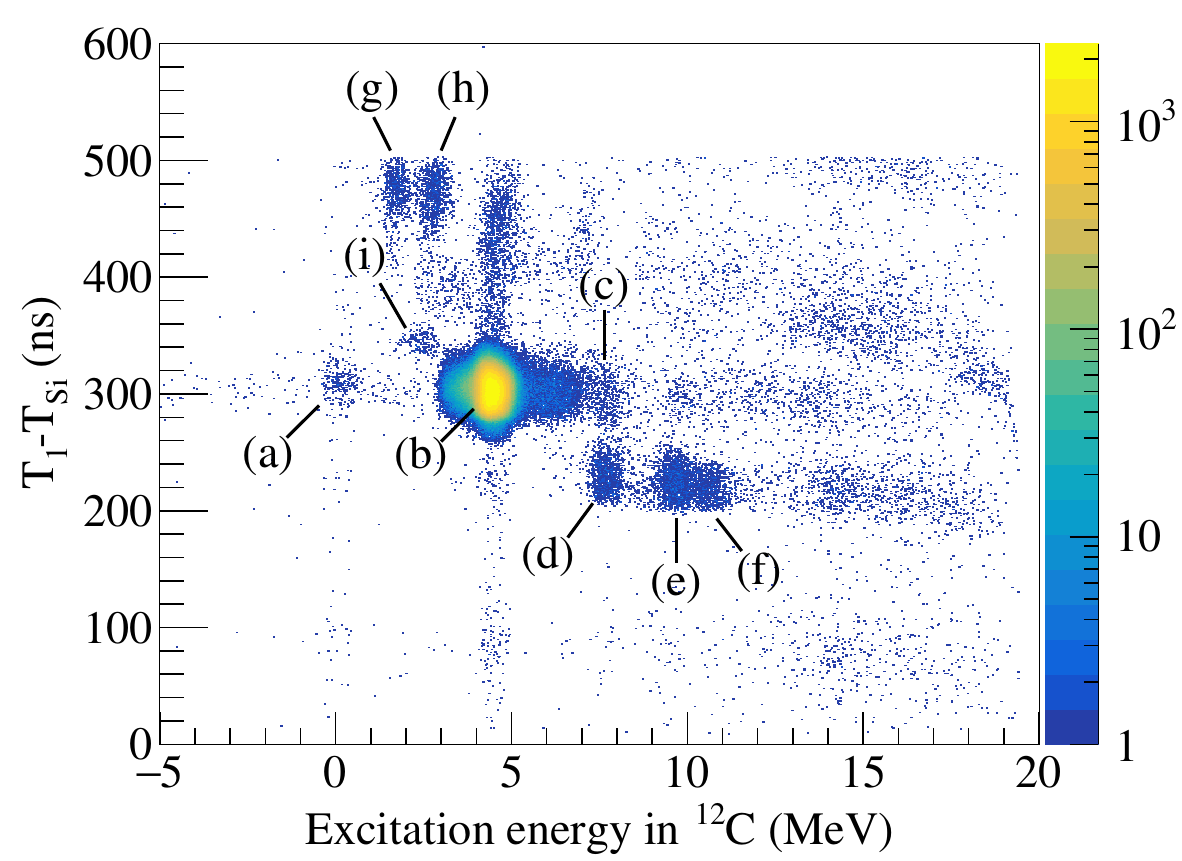}
\caption{Timing difference between the first PPAC and the DSSD ($T_{1}-T_{\rm Si}$) vs excitation energy in $^{12}{\rm C}$. See the text for details.}
\label{fig:T1Si}
\end{figure}

Figure \ref{fig:T1Si} shows the ($T_{1}-T_{\rm Si}$) against the excitation energy in $^{12}{\rm C}$ calculated from the angle and energy of the $\alpha$ particles in the DSSD. The observed groups of events in this two-dimensional (2D) scatter plot predominantly correspond to $^{12}{\rm C}$, which constitutes the primary component of the target. Groups (a), (b), and (c) in Fig. \ref{fig:T1Si} correspond to $^{12}{\rm C}$ ions originating from $^{12}{\rm C}+\alpha$ elastic scattering and inelastic scattering populating the $^{12}{\rm C}$(4.44) and $^{12}{\rm C}$(7.65) states, respectively. Groups (d), (e), and (f) represent the $\alpha$ particles resulting from the $\alpha$ decay of the $^{12}{\rm C}$(7.65), $^{12}{\rm C}$(9.64), and $^{12}{\rm C}$(10.847) states, respectively, as they exhibit shorter ToF compared to $^{12}{\rm C}$ ions. 

Groups (g) and (h) in Fig. \ref{fig:T1Si} indicate the presence of $^{16}{\rm O}$ contaminants in the target. Through examining the energy of the associated $\alpha$ particles, we have confirmed that groups (g) and (h) correspond to $^{16}{\rm O}$ ions resulting from $^{16}{\rm O}$+$\alpha$ inelastic scattering populating the $^{16}{\rm O}$(6.05, 6.13) and $^{16}{\rm O}$(6.92, 7.12) states, respectively. Two-body kinematics calculations reveal that the kinetic energies of the $^{16}{\rm O}$ ions for these two groups are approximately 0.90 and 0.88 MeV/nucleon, respectively. These energies are insufficient for $^{16}{\rm O}$(g.s.) ions to reach the second PPAC.

The inelastic scattering of $\alpha$ particles on the $^{16}{\rm O}$ contaminant also resulted in the population of $^{16}{\rm O}$ in excited states above the $\alpha$ decay threshold. Some of these states, with appropriate spin-parity, are open to $\alpha$ decay. Consequently, the production of $^{12}{\rm C}$(g.s.) ions can occur through the reaction $^{16}{\rm O}+\alpha\to$ $^{12}{\rm C}+\alpha+\alpha$. This process introduces a minor enhancement in the counts near the Hoyle state. However, as discussed in Sec. \ref{sec:spectrum}, this contribution can be effectively eliminated through proper fitting.

\begin{figure}[htbp] 
	\centering 
	\captionsetup{justification=RaggedRight}
	\begin{subfigure}[b]{0.48\textwidth}
	\includegraphics[width=\textwidth]{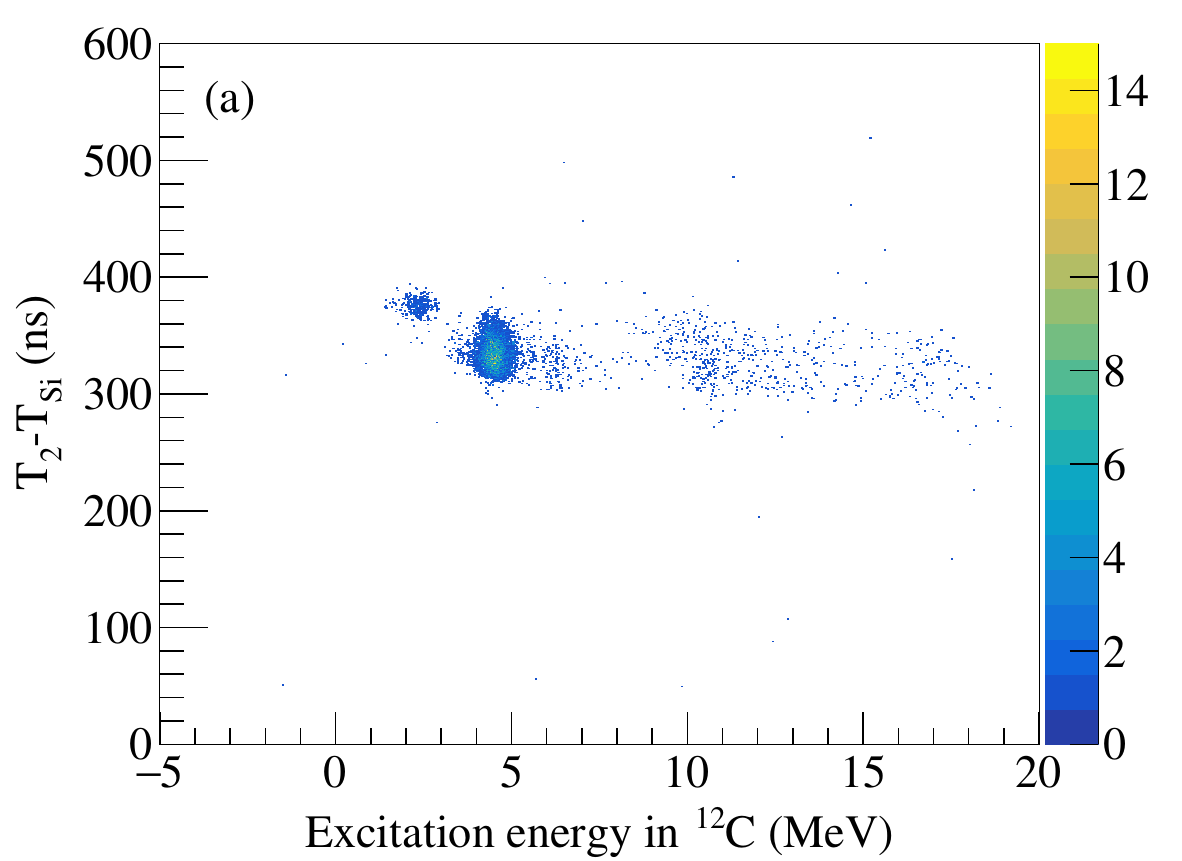}
	\captionlistentry{}
	\label{fig:T2TSi_13C}
	\end{subfigure}
	\hspace{2cm}
	\begin{subfigure}[b]{0.48\textwidth}
	\includegraphics[width=\textwidth]{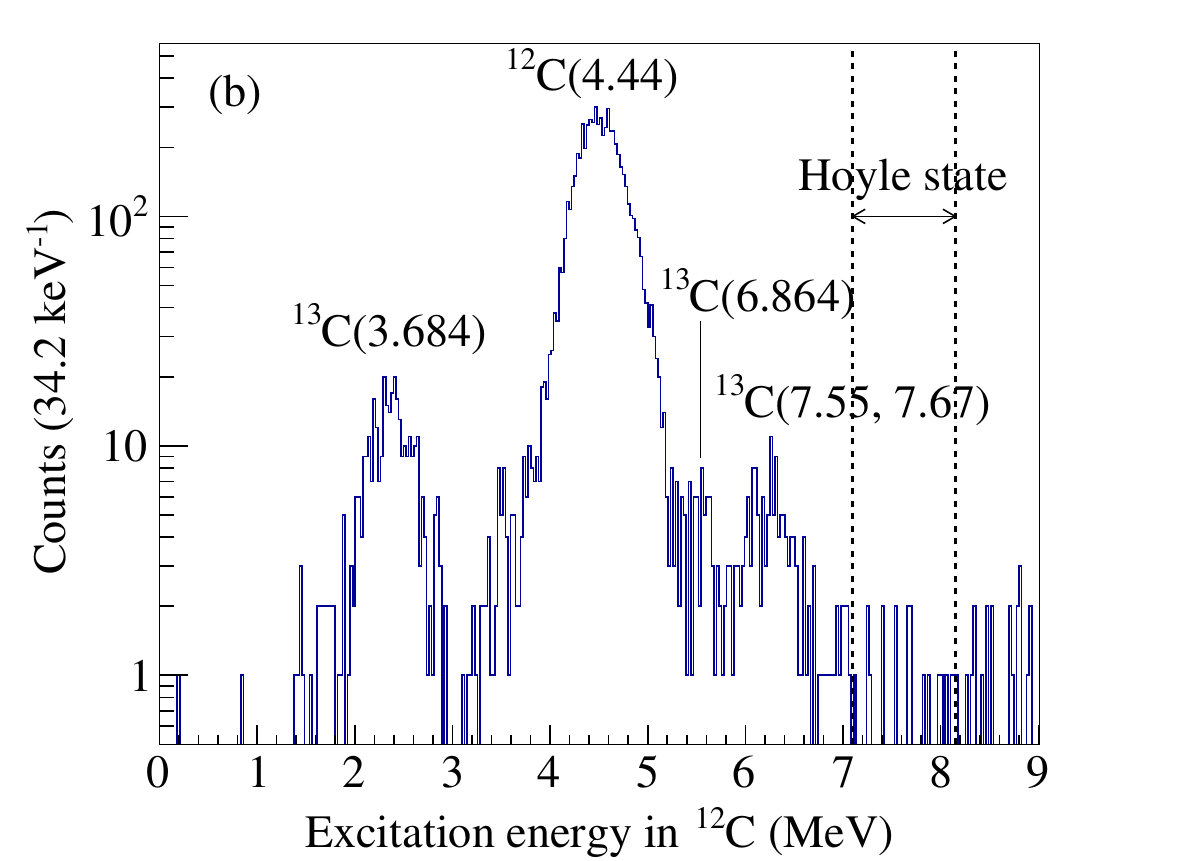}
	\captionlistentry{}
	\label{fig:excitation_13C}
	\end{subfigure}
	\caption{Measurement conducted with $^{13}{\rm C}$ target under the same experimental conditions as for the Hoyle state. (a) Plot of $T_{2}-T_{\rm Si}$ vs excitation energy for measurement conducted using the $^{13}{\rm C}$ target. (b) Excitation energy spectrum obtained from measurement using the $^{13}{\rm C}$ target. The labels adjacent to each peak indicate their corresponding origins. The black dashed lines outline the energy range for the Hoyle state. The excitation energy values in both figures were calculated using the mass of $^{12}{\rm C}$.}
	\label{fig:13C}
\end{figure}

While we used an isotopically enriched $^{12}{\rm C}$ target, the $^{13}{\rm C}$ isotope was still present. To assess the influence of $^{13}{\rm C}$, we conducted additional measurements using a 1-mg/cm$^{2}$ $^{13}{\rm C}$ target with the same experimental setup and magnetic rigidity as the measurements for the Hoyle state. Figure \ref{fig:T2TSi_13C} displays the 2D timing spectrum obtained from the measurement with $^{13}{\rm C}$ target, while Fig. \ref{fig:excitation_13C} presents the excitation-energy spectrum obtained by projecting the data in Fig. \ref{fig:T2TSi_13C} onto the $x$ axis.

The $^{13}{\rm C}$ target contains $^{12}{\rm C}$ at the level of few percent, resulting in a prominent peak originating from the $^{12}{\rm C}$(4.44) state. The group in Fig. \ref{fig:T2TSi_13C} characterized by an excitation energy $E_{x}$=2.4 MeV and $T_{2}-T_{\rm Si}$=380 ns corresponds to $^{13}{\rm C}$(g.s.) ions generated through $^{13}{\rm C}+\alpha$ inelastic scattering, specifically populating the $^{13}{\rm C}$(3.684) state. The group (i) observed in Fig. \ref{fig:T1Si} is attributed to the same origin. 

By comparing the ratio of counts in group (i) to the counts in group (b) obtained from the $^{13}{\rm C}$ target measurements with those from the Hoyle state measurements, we were able to make an estimation of the fraction of $^{13}{\rm C}$ in the enriched $^{12}{\rm C}$ target. Assuming that the $^{13}{\rm C}$ target is composed entirely of $^{13}{\rm C}$, our analysis yielded a fraction of 0.17\% for the $^{13}{\rm C}$ contaminant in the $^{12}{\rm C}$-enriched target. However, it is only an upper limit because the enrichment in the $^{13}{\rm C}$ target is not 100\%.

Furthermore, the observed groups with $E_{x}$=5.4 and 6.3 MeV correspond to $^{12}{\rm C}$ ions originating from $^{13}{\rm C}+\alpha\to$ $^{12}{\rm C}+n+\alpha$ reactions. Initially, $^{13}{\rm C}+\alpha$ inelastic scattering populates the $^{13}{\rm C}$(6.864), $^{13}{\rm C}$(7.55), and $^{13}{\rm C}$(7.67) states, which subsequently decay to a $^{12}{\rm C}$(g.s.) ion and a neutron. Figure \ref{fig:excitation_13C} illustrates the energy spectrum, where we observe some counts in the energy region of the Hoyle state. These additional counts could potentially enhance the signal for the Hoyle state. However, we can eliminate this contribution through a proper scaling procedure, which will be discussed in detail in Sec. \ref{sec:spectrum}.

\subsection{Excitation energy spectra}\label{sec:spectrum}
\begin{figure}[htbp] 
	\centering 
	\captionsetup{justification=RaggedRight}
	\includegraphics[width=0.48\textwidth]{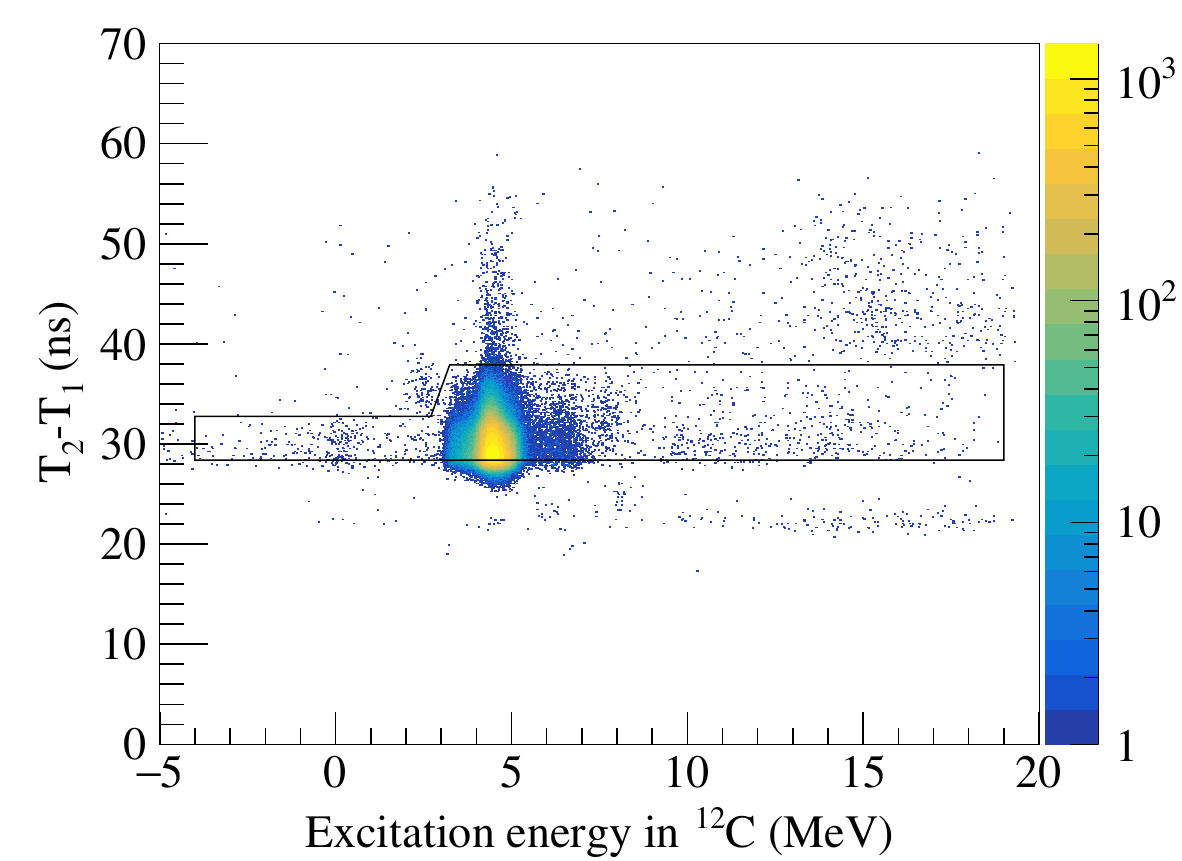}
	\caption{ToF between the two PPACs vs excitation energy in $^{12}{\rm C}$. The black polygon represents the cut applied to select $^{12}{\rm C}$ events.}
	\label{fig:tof}
\end{figure}

Figure \ref{fig:tof} displays the ToF measurements for particles traveling from the first to the second PPAC of the TexPPACS. This 2D ToF spectrum offers improved separation between $^{12}{\rm C}$ and $\alpha$ particles compared to Fig. \ref{fig:T1Si}. By applying a polygon cut for $^{12}{\rm C}$ as indicated in this figure and projecting the data onto the $x$ axis, we obtained the excitation-energy spectrum for the coincidence events.

\begin{figure}[htbp] 
	\centering 
	\captionsetup{justification=RaggedRight}
	\includegraphics[width=0.48\textwidth]{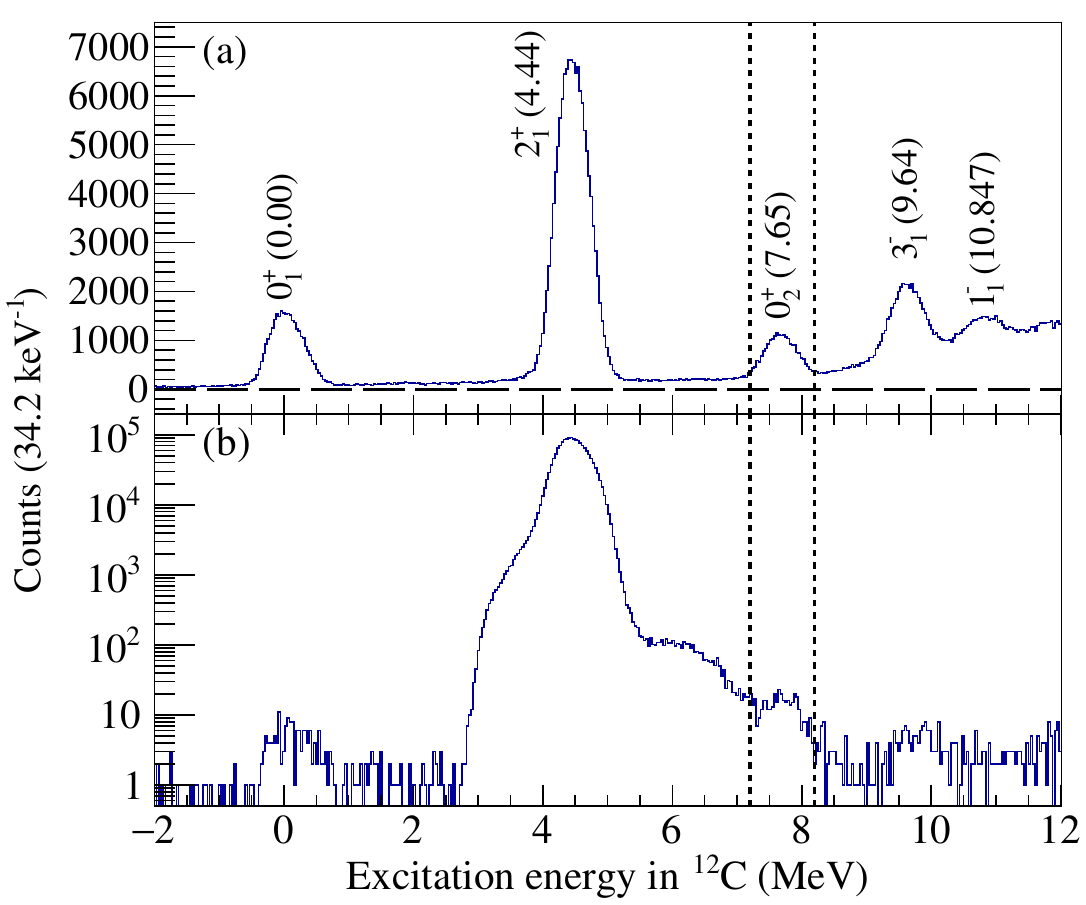}
	\caption{Excitation-energy spectra of $^{12}{\rm C}$ for (a) the singles events and (b) the coincidence with $^{12}{\rm C}$ events in the inelastic $\alpha$ scattering.}
	\label{fig:excitation}
\end{figure}

Figure \ref{fig:excitation} displays the excitation-energy spectra of (a) singles and (b) coincidence with $^{12}{\rm C}$ recoil events. The dashed vertical lines identify the location of the Hoyle state. It is clearly visible in the singles and the $\alpha$+$^{12}{\rm C}$ coincidence spectra. However, it is important to note that there are three sources of contamination affecting the Hoyle state peak: a ``shoulder'' on the left side of the peak, as well as enhancements originating from $^{16}{\rm O}$ and $^{13}{\rm C}$. The subsequent discussion will address the methods employed to eliminate their influences.

The broad feature observed to the left of the Hoyle peak in the coincidence spectrum is attributed to the edge effect of the DSSD silicon detector. Each strip on the DSSD has a width of 3000 $\mu$m, with two 100-$\mu$m-wide dead regions on either side. The probability of inelastically scattered $\alpha$ particles hitting the overlapping area between the front and back dead regions is $(100/3000)^2 \approx 0.1\%$. Consequently, the charges induced by these $\alpha$ particles were not fully collected. As a result, slightly higher excitation energies in $^{12}{\rm C}$ were obtained, leading to the appearance of the ``shoulder'' feature. This hypothesis was verified by modifying the energy matching conditions for the front and back strips of the DSSD. We observed a significant drop in the ``shoulder'' feature when we restricted the condition on the energy difference between the front and back strips, while the intensities of other peaks remained less changed. In our analysis, we used an energy matching condition of 0.5\%, which is consistent with the energy resolution of the DSSD, i.e., $|E_{\rm front}-E_{\rm back}|/E_{\rm front}<0.5\%$.

\begin{figure}[htbp] 
	\centering 
	\captionsetup{justification=RaggedRight}
	\includegraphics[width=0.48\textwidth]{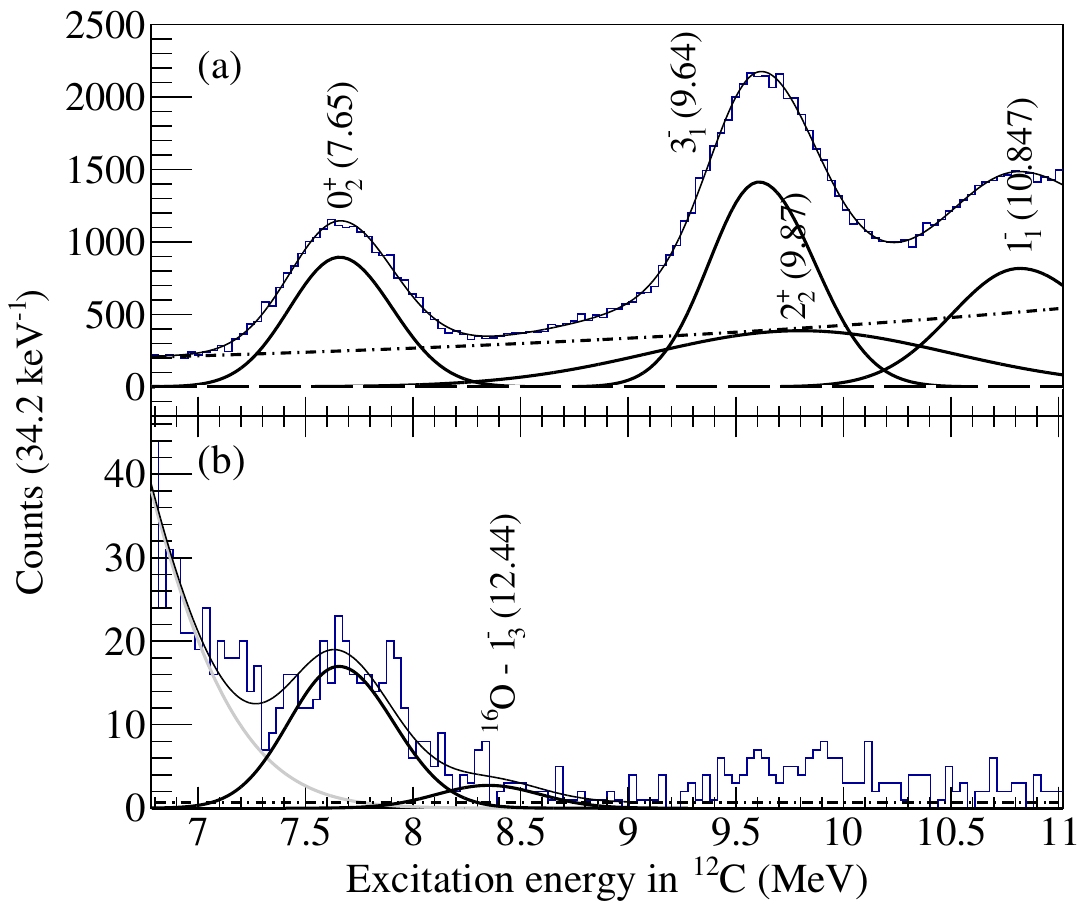}
	\caption{Excitation-energy spectra of $^{12}{\rm C}$ around the Hoyle state for (a) the singles events and (b) the coincidence events in the inelastic $\alpha$ scattering. The thick solid lines represent the fitted Gaussian functions for the $0^{+}_{2}$, $3^{-}_{1}$, $2^{+}_{2}$, $1^{-}_{1}$ states in $^{12}{\rm C}$ and the $1^{-}_{3}$ state in $^{16}{\rm O}$. The states are labeled near each peak. The gray solid line corresponds to the fitted Gaussian function for the shoulder around $E_{x}=6$ MeV. The black dash-dotted lines represent the background, and the thin solid line represents the sum of these functions.}
	\label{fig:excitation_fitted}
\end{figure}

To determine the yields of singles and coincidence events, both spectra were fitted using Gaussian functions for the Hoyle and $3^{-}_{1}$ states, as well as other peaks, while a smooth function was employed for the continuum. The fitted results are shown in Fig. \ref{fig:excitation_fitted}. The centroids and widths of the Gaussian functions were adjusted to reproduce the singles spectrum, and the same parameters were utilized for the coincidence spectrum. The continuum was fitted by an exponential function, while two other functions were also tested to estimate the systematic uncertainty: a semiphenomenological function obtained from Ref. \cite{Semi} with an added constant offset, and a linear function. The measured spectra were then subtracted by the fit functions for the continuum, and the remaining spectra were integrated to obtain the yields of the Hoyle state. This approach was employed to mitigate errors resulting from discrepancies between the Gaussian fit function and the actual measured peak shape.

The $^{16}{\rm O}+\alpha$ inelastic scattering can enhance the counts in the region of the Hoyle state. Specifically, the $1_{3}^{-}$ state in $^{16}{\rm O}$ with excitation energy of 12.44 MeV was observed in the coincidence spectrum shown in Fig. \ref{fig:excitation_fitted}. To obtain an accurate count of the Hoyle state, this peak was fitted using a Gaussian function with the same width parameter as the Hoyle state (determined by the experimental energy resolution).

The coincident yields within the energy range of the Hoyle state from the $^{13}{\rm C}$ target measurement and the $^{12}{\rm C}$-enriched target measurement were normalized by scaling them with the ratio of the counts of the $^{13}{\rm C}$(3.684) peaks observed in both measurements. This normalization procedure allows us to determine the additional counts in the region of the Hoyle state originating from the $^{13}{\rm C}$ contaminants. The contribution from the $^{13}{\rm C}$ contaminants in the $E_{x}=7.65$ MeV peak is estimated to be 3\%, and it was subtracted from the total counts of the Hoyle state peak.

The numbers of actual counts under the Hoyle state peaks in the singles events and coincidence events were determined to be $N_{\rm{Scaled}}=(1.570\pm0.037)\times10^{4}$ and $N_{\rm{Coinc}}=291.2\pm20.7$, respectively. According to Eq. \ref{ratio-1}, the radiative branching ratio was determined to be $(\Gamma_{\rm{rad}}/\Gamma)_{\rm expo}\times10^{4}=3.94\pm0.36({\rm stat.})\pm0.13({\rm syst.})$. Here the systematic uncertainty arose from the uncertainty of the MDM-TexPPACS efficiency. The semiphenomenological continuum gave $(\Gamma_{\rm{rad}}/\Gamma)_{\rm semi}\times10^{4}=3.92\pm0.34({\rm stat.})\pm0.13({\rm syst.})$. We also fitted the background using a linear function from $E_x=6.9 - 8.4$ MeV as the extreme assumption, and this gave $(\Gamma_{\rm{rad}}/\Gamma)_{\rm line}\times10^{4}=3.95\pm0.37({\rm stat.})\pm0.13({\rm syst.})$. The largest difference between the three yields was considered as an addition to the systematic uncertainty arising from the ambiguity of the continuum function. Consequently, the present radiative branching ratio of the Hoyle state in $^{12}{\rm C}$ was determined to be $\Gamma_{\rm{rad}}/\Gamma\times10^{4}=3.94\pm0.36({\rm stat.})\pm0.16({\rm syst.})$.

Another measurement was conducted with an increase of 0.5\% in the magnetic field strength. The analysis procedure followed the same methodology as described above. The obtained radiative branching ratio was found to be $\Gamma_{\rm{rad}}/\Gamma\times10^{4}=4.07\pm0.51({\rm stat.})\pm0.16({\rm syst.})$, which is in agreement with the previous result within uncertainty. This additional measurement at slightly different magnetic rigidity serves as a confirmation of the reliability of the analysis procedures and the independence of the result to a specific choice of magnetic rigidity within the allowed range. The second measurement was combined with the first to reduce the statistical uncertainty.    

\begin{table}[htbp]
	\caption{Quantities used to evaluate $\Gamma_{\rm{rad}}/\Gamma$\label{tab:information}}
	\begin{ruledtabular}
	  \begin{tabular}{lll}
		Quantity & Measurement 1 & Measurement 2\\
		\midrule
		$N_{\rm{Coinc}}$ & $291.2\pm20.7$ & $150.6\pm16.5$\\
		$N_{\rm{Scaled}}$ & $(1.570\pm0.037)\times10^{4}$ & $(7.789\pm0.243)\times10^{3}$\\
		$\epsilon$\footnote{The uncertainty in $\epsilon$ is systematic, while all other uncertainties are statistical.} & $0.95\pm0.03$ & $0.96\pm0.03$ \\
		$F_{5+}$ & \multicolumn{2}{c}{$0.495\pm0.026$} \\
		$\Gamma_{\rm{rad}}/\Gamma$ & $(3.94\pm0.36)\times10^{4}$ & $(4.07\pm0.51)\times10^{4}$\\
	  \end{tabular}
	\end{ruledtabular}
\end{table}

Table \ref{tab:information} summarizes the quantities used to evaluate radiative branching ratio from the two measurements. Considering both measurements, we report the radiative branching ratio of the Hoyle state in $^{12}{\rm C}$ to be $\Gamma_{\rm{rad}}/\Gamma\times10^{4}=4.0\pm0.3({\rm stat.})\pm0.16({\rm syst.})$. This result is the statistical uncertainty-weighted mean value obtained from the two measurements.

\section{Conclusion}
Table \ref{tab:comparison} presents a comparison of the result obtained in this study with other recent measurements and the adopted value. The present result is in good agreement with the previous measurements, except for the Kib\'edi \cite{Kibedi2020} measurement which deviates significantly. The adopted value was determined based on the weighted mean of the measurements conducted before 1976. Tsumura \cite{Tsumura2021} performed a measurement with magnetic selection, but their result exhibits a large relative uncertainty of approximately 18.6\%. Our present work shows a substantial improvement achieving approximately 11.5\%, which included the systematic uncertainty of 4\%. These results also complement the findings from an alternative experimental probe, namely the $\beta$ decay of $^{12}{\rm B}$ and $^{12}{\rm N}$ \cite{fynbo-2009,fynbo-2016}. Considering both the older and recent results, it may be appropriate to exclude the $\gamma$-particle spectroscopy result reported by Kib\'edi \cite{Kibedi2020} from further consideration.

\begin{table}[htbp]
	\captionsetup{justification=RaggedRight}
	\caption{Comparison of the recent experimental results.\label{tab:comparison}}
	\begin{ruledtabular}
	  \begin{tabular}{lcl}
		Reference & $\Gamma_{\rm{rad}}/\Gamma\times10^{4}$ & Methodology\\
		\midrule
		Kelley \cite{KELLEY201771} & $4.16\pm0.11$ & Adopted value\\
		Kib\'edi \cite{Kibedi2020} & $6.2\pm0.6$\footnote{Systematic uncertainty is not reported.} & $\gamma$-particle spectroscopy\\
		Tsumura \cite{Tsumura2021} & $4.3\pm0.8$\footnote{Uncertainty is large because systematic uncertainty is included.} & Charged-particle spectroscopy \\
		This work & $4.0\pm0.3$\footnote{Systematic uncertainty (4.0\%) is not included.} & Charged-particle spectroscopy\\
	  \end{tabular}
	\end{ruledtabular}
\end{table}

\begin{acknowledgments}
The authors thank the operation team at Cyclotron Institute, Texas A\&M University, for the reliable operation of the facilities. The authors also thank Dr. Heshani Jayatissa and Dr. Takahiro Kawabata for their helpful discussions. This work was supported by the U.S. Department of Energy, Office of Science, Office of Nuclear Science under Award No. DE-FG02-93ER40773, and by the National Nuclear Security Administration through the Center for Excellence in Nuclear Training and University-Based Research (CENTAUR) under Grant No. DE-NA0003841.
\end{acknowledgments}

\bibliography{HoylePaper}

\begin{thebibliography}{19}%
\makeatletter
\providecommand \@ifxundefined [1]{%
 \@ifx{#1\undefined}
}%
\providecommand \@ifnum [1]{%
 \ifnum #1\expandafter \@firstoftwo
 \else \expandafter \@secondoftwo
 \fi
}%
\providecommand \@ifx [1]{%
 \ifx #1\expandafter \@firstoftwo
 \else \expandafter \@secondoftwo
 \fi
}%
\providecommand \natexlab [1]{#1}%
\providecommand \enquote  [1]{``#1''}%
\providecommand \bibnamefont  [1]{#1}%
\providecommand \bibfnamefont [1]{#1}%
\providecommand \citenamefont [1]{#1}%
\providecommand \href@noop [0]{\@secondoftwo}%
\providecommand \href [0]{\begingroup \@sanitize@url \@href}%
\providecommand \@href[1]{\@@startlink{#1}\@@href}%
\providecommand \@@href[1]{\endgroup#1\@@endlink}%
\providecommand \@sanitize@url [0]{\catcode `\\12\catcode `\$12\catcode `\&12\catcode `\#12\catcode `\^12\catcode `\_12\catcode `\%12\relax}%
\providecommand \@@startlink[1]{}%
\providecommand \@@endlink[0]{}%
\providecommand \url  [0]{\begingroup\@sanitize@url \@url }%
\providecommand \@url [1]{\endgroup\@href {#1}{\urlprefix }}%
\providecommand \urlprefix  [0]{URL }%
\providecommand \Eprint [0]{\href }%
\providecommand \doibase [0]{https://doi.org/}%
\providecommand \selectlanguage [0]{\@gobble}%
\providecommand \bibinfo  [0]{\@secondoftwo}%
\providecommand \bibfield  [0]{\@secondoftwo}%
\providecommand \translation [1]{[#1]}%
\providecommand \BibitemOpen [0]{}%
\providecommand \bibitemStop [0]{}%
\providecommand \bibitemNoStop [0]{.\EOS\space}%
\providecommand \EOS [0]{\spacefactor3000\relax}%
\providecommand \BibitemShut  [1]{\csname bibitem#1\endcsname}%
\let\auto@bib@innerbib\@empty
\bibitem [{\citenamefont {Hoyle}(1954)}]{hoyle1954nuclear}%
  \BibitemOpen
  \bibfield  {author} {\bibinfo {author} {\bibfnamefont {F.}~\bibnamefont {Hoyle}},\ }\href {https://ui.adsabs.harvard.edu/link_gateway/1954ApJS....1..121H/doi:10.1086/190005} {\bibfield  {journal} {\bibinfo  {journal} {Astrophys. J. Suppl., vol. 1, p. 121 (1954)}\ }\textbf {\bibinfo {volume} {1}},\ \bibinfo {pages} {121} (\bibinfo {year} {1954})}\BibitemShut {NoStop}%
\bibitem [{\citenamefont {Kib\'edi}\ \emph {et~al.}(2020)\citenamefont {Kib\'edi}, \citenamefont {Alshahrani}, \citenamefont {Stuchbery}, \citenamefont {Larsen}, \citenamefont {G\"orgen}, \citenamefont {Siem}, \citenamefont {Guttormsen}, \citenamefont {Giacoppo}, \citenamefont {Morales}, \citenamefont {Sahin}, \citenamefont {Tveten}, \citenamefont {Garrote}, \citenamefont {Campo}, \citenamefont {Eriksen}, \citenamefont {Klintefjord} \emph {et~al.}}]{Kibedi2020}%
  \BibitemOpen
  \bibfield  {author} {\bibinfo {author} {\bibfnamefont {T.}~\bibnamefont {Kib\'edi}}, \bibinfo {author} {\bibfnamefont {B.}~\bibnamefont {Alshahrani}}, \bibinfo {author} {\bibfnamefont {A.~E.}\ \bibnamefont {Stuchbery}}, \bibinfo {author} {\bibfnamefont {A.~C.}\ \bibnamefont {Larsen}}, \bibinfo {author} {\bibfnamefont {A.}~\bibnamefont {G\"orgen}}, \bibinfo {author} {\bibfnamefont {S.}~\bibnamefont {Siem}}, \bibinfo {author} {\bibfnamefont {M.}~\bibnamefont {Guttormsen}}, \bibinfo {author} {\bibfnamefont {F.}~\bibnamefont {Giacoppo}}, \bibinfo {author} {\bibfnamefont {A.~I.}\ \bibnamefont {Morales}}, \bibinfo {author} {\bibfnamefont {E.}~\bibnamefont {Sahin}}, \bibinfo {author} {\bibfnamefont {G.~M.}\ \bibnamefont {Tveten}}, \bibinfo {author} {\bibfnamefont {F.~L.~B.}\ \bibnamefont {Garrote}}, \bibinfo {author} {\bibfnamefont {L.~C.}\ \bibnamefont {Campo}}, \bibinfo {author} {\bibfnamefont {T.~K.}\ \bibnamefont {Eriksen}}, \bibinfo {author} {\bibfnamefont {M.}~\bibnamefont {Klintefjord}}, \emph {et~al.},\ }\href {https://doi.org/10.1103/PhysRevLett.125.182701} {\bibfield  {journal} {\bibinfo  {journal} {Phys. Rev. Lett.}\ }\textbf {\bibinfo {volume} {125}},\ \bibinfo {pages} {182701} (\bibinfo {year} {2020})}\BibitemShut {NoStop}%
\bibitem [{\citenamefont {Alburger}(1961)}]{Ref1961}%
  \BibitemOpen
  \bibfield  {author} {\bibinfo {author} {\bibfnamefont {D.~E.}\ \bibnamefont {Alburger}},\ }\href {https://doi.org/10.1103/PhysRev.124.193} {\bibfield  {journal} {\bibinfo  {journal} {Phys. Rev.}\ }\textbf {\bibinfo {volume} {124}},\ \bibinfo {pages} {193} (\bibinfo {year} {1961})}\BibitemShut {NoStop}%
\bibitem [{\citenamefont {Seeger}\ and\ \citenamefont {Kavanagh}(1963)}]{Ref1963-1}%
  \BibitemOpen
  \bibfield  {author} {\bibinfo {author} {\bibfnamefont {P.}~\bibnamefont {Seeger}}\ and\ \bibinfo {author} {\bibfnamefont {R.}~\bibnamefont {Kavanagh}},\ }\href {https://doi.org/https://doi.org/10.1016/0029-5582(63)90630-8} {\bibfield  {journal} {\bibinfo  {journal} {Nucl. Phys.}\ }\textbf {\bibinfo {volume} {46}},\ \bibinfo {pages} {577} (\bibinfo {year} {1963})}\BibitemShut {NoStop}%
\bibitem [{\citenamefont {Hall}\ and\ \citenamefont {Tanner}(1964)}]{Ref1963-2}%
  \BibitemOpen
  \bibfield  {author} {\bibinfo {author} {\bibfnamefont {I.}~\bibnamefont {Hall}}\ and\ \bibinfo {author} {\bibfnamefont {N.}~\bibnamefont {Tanner}},\ }\href {https://doi.org/https://doi.org/10.1016/0029-5582(64)90646-7} {\bibfield  {journal} {\bibinfo  {journal} {Nucl. Phys.}\ }\textbf {\bibinfo {volume} {53}},\ \bibinfo {pages} {673} (\bibinfo {year} {1964})}\BibitemShut {NoStop}%
\bibitem [{\citenamefont {Chamberlin}\ \emph {et~al.}(1974)\citenamefont {Chamberlin}, \citenamefont {Bodansky}, \citenamefont {Jacobs},\ and\ \citenamefont {Oberg}}]{Ref1974}%
  \BibitemOpen
  \bibfield  {author} {\bibinfo {author} {\bibfnamefont {D.}~\bibnamefont {Chamberlin}}, \bibinfo {author} {\bibfnamefont {D.}~\bibnamefont {Bodansky}}, \bibinfo {author} {\bibfnamefont {W.~W.}\ \bibnamefont {Jacobs}},\ and\ \bibinfo {author} {\bibfnamefont {D.~L.}\ \bibnamefont {Oberg}},\ }\href {https://doi.org/10.1103/PhysRevC.9.69} {\bibfield  {journal} {\bibinfo  {journal} {Phys. Rev. C}\ }\textbf {\bibinfo {volume} {9}},\ \bibinfo {pages} {69} (\bibinfo {year} {1974})}\BibitemShut {NoStop}%
\bibitem [{\citenamefont {Davids}\ \emph {et~al.}(1975)\citenamefont {Davids}, \citenamefont {Pardo},\ and\ \citenamefont {Obst}}]{Ref1975-1}%
  \BibitemOpen
  \bibfield  {author} {\bibinfo {author} {\bibfnamefont {C.~N.}\ \bibnamefont {Davids}}, \bibinfo {author} {\bibfnamefont {R.~C.}\ \bibnamefont {Pardo}},\ and\ \bibinfo {author} {\bibfnamefont {A.~W.}\ \bibnamefont {Obst}},\ }\href {https://doi.org/10.1103/PhysRevC.11.2063} {\bibfield  {journal} {\bibinfo  {journal} {Phys. Rev. C}\ }\textbf {\bibinfo {volume} {11}},\ \bibinfo {pages} {2063} (\bibinfo {year} {1975})}\BibitemShut {NoStop}%
\bibitem [{\citenamefont {Mak}\ \emph {et~al.}(1975)\citenamefont {Mak}, \citenamefont {Evans}, \citenamefont {Ewan}, \citenamefont {McDonald},\ and\ \citenamefont {Alexander}}]{Ref1975-2}%
  \BibitemOpen
  \bibfield  {author} {\bibinfo {author} {\bibfnamefont {H.~B.}\ \bibnamefont {Mak}}, \bibinfo {author} {\bibfnamefont {H.~C.}\ \bibnamefont {Evans}}, \bibinfo {author} {\bibfnamefont {G.~T.}\ \bibnamefont {Ewan}}, \bibinfo {author} {\bibfnamefont {A.~B.}\ \bibnamefont {McDonald}},\ and\ \bibinfo {author} {\bibfnamefont {T.~K.}\ \bibnamefont {Alexander}},\ }\href {https://doi.org/10.1103/PhysRevC.12.1158} {\bibfield  {journal} {\bibinfo  {journal} {Phys. Rev. C}\ }\textbf {\bibinfo {volume} {12}},\ \bibinfo {pages} {1158} (\bibinfo {year} {1975})}\BibitemShut {NoStop}%
\bibitem [{\citenamefont {Markham}\ \emph {et~al.}(1976)\citenamefont {Markham}, \citenamefont {Austin},\ and\ \citenamefont {Shahabuddin}}]{Ref1976-1}%
  \BibitemOpen
  \bibfield  {author} {\bibinfo {author} {\bibfnamefont {R.}~\bibnamefont {Markham}}, \bibinfo {author} {\bibfnamefont {S.~M.}\ \bibnamefont {Austin}},\ and\ \bibinfo {author} {\bibfnamefont {M.}~\bibnamefont {Shahabuddin}},\ }\href {https://doi.org/https://doi.org/10.1016/0375-9474(76)90458-9} {\bibfield  {journal} {\bibinfo  {journal} {Nucl. Phys. A}\ }\textbf {\bibinfo {volume} {270}},\ \bibinfo {pages} {489} (\bibinfo {year} {1976})}\BibitemShut {NoStop}%
\bibitem [{\citenamefont {Obst}\ and\ \citenamefont {Braithwaite}(1976)}]{Ref1976-2}%
  \BibitemOpen
  \bibfield  {author} {\bibinfo {author} {\bibfnamefont {A.~W.}\ \bibnamefont {Obst}}\ and\ \bibinfo {author} {\bibfnamefont {W.~J.}\ \bibnamefont {Braithwaite}},\ }\href {https://doi.org/10.1103/PhysRevC.13.2033} {\bibfield  {journal} {\bibinfo  {journal} {Phys. Rev. C}\ }\textbf {\bibinfo {volume} {13}},\ \bibinfo {pages} {2033} (\bibinfo {year} {1976})}\BibitemShut {NoStop}%
\bibitem [{\citenamefont {Kelley}\ \emph {et~al.}(2017)\citenamefont {Kelley}, \citenamefont {Purcell},\ and\ \citenamefont {Sheu}}]{KELLEY201771}%
  \BibitemOpen
  \bibfield  {author} {\bibinfo {author} {\bibfnamefont {J.}~\bibnamefont {Kelley}}, \bibinfo {author} {\bibfnamefont {J.}~\bibnamefont {Purcell}},\ and\ \bibinfo {author} {\bibfnamefont {C.}~\bibnamefont {Sheu}},\ }\href {https://doi.org/https://doi.org/10.1016/j.nuclphysa.2017.07.015} {\bibfield  {journal} {\bibinfo  {journal} {Nucl. Phys. A}\ }\textbf {\bibinfo {volume} {968}},\ \bibinfo {pages} {71} (\bibinfo {year} {2017})}\BibitemShut {NoStop}%
\bibitem [{\citenamefont {Tsumura}\ \emph {et~al.}(2021)\citenamefont {Tsumura}, \citenamefont {Kawabata}, \citenamefont {Takahashi}, \citenamefont {Adachi}, \citenamefont {Akimune}, \citenamefont {Ashikaga}, \citenamefont {Baba}, \citenamefont {Fujikawa}, \citenamefont {Fujimura}, \citenamefont {Fujioka}, \citenamefont {Furuno}, \citenamefont {Hashimoto}, \citenamefont {Harada}, \citenamefont {Ichikawa}, \citenamefont {Inaba} \emph {et~al.}}]{Tsumura2021}%
  \BibitemOpen
  \bibfield  {author} {\bibinfo {author} {\bibfnamefont {M.}~\bibnamefont {Tsumura}}, \bibinfo {author} {\bibfnamefont {T.}~\bibnamefont {Kawabata}}, \bibinfo {author} {\bibfnamefont {Y.}~\bibnamefont {Takahashi}}, \bibinfo {author} {\bibfnamefont {S.}~\bibnamefont {Adachi}}, \bibinfo {author} {\bibfnamefont {H.}~\bibnamefont {Akimune}}, \bibinfo {author} {\bibfnamefont {S.}~\bibnamefont {Ashikaga}}, \bibinfo {author} {\bibfnamefont {T.}~\bibnamefont {Baba}}, \bibinfo {author} {\bibfnamefont {Y.}~\bibnamefont {Fujikawa}}, \bibinfo {author} {\bibfnamefont {H.}~\bibnamefont {Fujimura}}, \bibinfo {author} {\bibfnamefont {H.}~\bibnamefont {Fujioka}}, \bibinfo {author} {\bibfnamefont {T.}~\bibnamefont {Furuno}}, \bibinfo {author} {\bibfnamefont {T.}~\bibnamefont {Hashimoto}}, \bibinfo {author} {\bibfnamefont {T.}~\bibnamefont {Harada}}, \bibinfo {author} {\bibfnamefont {M.}~\bibnamefont {Ichikawa}}, \bibinfo {author} {\bibfnamefont {K.}~\bibnamefont {Inaba}}, \emph {et~al.},\ }\href {https://doi.org/https://doi.org/10.1016/j.physletb.2021.136283} {\bibfield  {journal} {\bibinfo  {journal} {Phys. Lett. B}\ }\textbf {\bibinfo {volume} {817}},\ \bibinfo {pages} {136283} (\bibinfo {year} {2021})}\BibitemShut {NoStop}%
\bibitem [{\citenamefont {{Micron Semiconductor Ltd.}}()}]{Micron}%
  \BibitemOpen
  \bibfield  {author} {\bibinfo {author} {\bibnamefont {{Micron Semiconductor Ltd.}}},\ }\href@noop {} {\bibinfo {title} {W1 double-sided silicon strip detector. \url{http://www.micronsemiconductor.co.uk/wp-content/uploads/2018/03/2018-Micron-Semiconductor-Ltd-Silicon-Catalogue_Long-Form.pdf}}}\BibitemShut {NoStop}%
\bibitem [{\citenamefont {Pringle}\ \emph {et~al.}(1986)\citenamefont {Pringle}, \citenamefont {Catford}, \citenamefont {Winfield}, \citenamefont {Lewis}, \citenamefont {Jelley}, \citenamefont {Allen},\ and\ \citenamefont {Coupland}}]{MDM}%
  \BibitemOpen
  \bibfield  {author} {\bibinfo {author} {\bibfnamefont {D.}~\bibnamefont {Pringle}}, \bibinfo {author} {\bibfnamefont {W.}~\bibnamefont {Catford}}, \bibinfo {author} {\bibfnamefont {J.}~\bibnamefont {Winfield}}, \bibinfo {author} {\bibfnamefont {D.}~\bibnamefont {Lewis}}, \bibinfo {author} {\bibfnamefont {N.}~\bibnamefont {Jelley}}, \bibinfo {author} {\bibfnamefont {K.}~\bibnamefont {Allen}},\ and\ \bibinfo {author} {\bibfnamefont {J.}~\bibnamefont {Coupland}},\ }\href {https://doi.org/https://doi.org/10.1016/0168-9002(86)91256-8} {\bibfield  {journal} {\bibinfo  {journal} {Nucl. Instrum. Methods Phys. Res., Sect. A}\ }\textbf {\bibinfo {volume} {245}},\ \bibinfo {pages} {230} (\bibinfo {year} {1986})}\BibitemShut {NoStop}%
\bibitem [{\citenamefont {Agostinelli}\ \emph {et~al.}(2003)\citenamefont {Agostinelli}, \citenamefont {Allison}, \citenamefont {Amako}, \citenamefont {Apostolakis}, \citenamefont {Araujo}, \citenamefont {Arce}, \citenamefont {Asai}, \citenamefont {Axen}, \citenamefont {Banerjee}, \citenamefont {Barrand}, \citenamefont {Behner}, \citenamefont {Bellagamba}, \citenamefont {Boudreau}, \citenamefont {Broglia}, \citenamefont {Brunengo} \emph {et~al.}}]{Geant4}%
  \BibitemOpen
  \bibfield  {author} {\bibinfo {author} {\bibfnamefont {S.}~\bibnamefont {Agostinelli}}, \bibinfo {author} {\bibfnamefont {J.}~\bibnamefont {Allison}}, \bibinfo {author} {\bibfnamefont {K.}~\bibnamefont {Amako}}, \bibinfo {author} {\bibfnamefont {J.}~\bibnamefont {Apostolakis}}, \bibinfo {author} {\bibfnamefont {H.}~\bibnamefont {Araujo}}, \bibinfo {author} {\bibfnamefont {P.}~\bibnamefont {Arce}}, \bibinfo {author} {\bibfnamefont {M.}~\bibnamefont {Asai}}, \bibinfo {author} {\bibfnamefont {D.}~\bibnamefont {Axen}}, \bibinfo {author} {\bibfnamefont {S.}~\bibnamefont {Banerjee}}, \bibinfo {author} {\bibfnamefont {G.}~\bibnamefont {Barrand}}, \bibinfo {author} {\bibfnamefont {F.}~\bibnamefont {Behner}}, \bibinfo {author} {\bibfnamefont {L.}~\bibnamefont {Bellagamba}}, \bibinfo {author} {\bibfnamefont {J.}~\bibnamefont {Boudreau}}, \bibinfo {author} {\bibfnamefont {L.}~\bibnamefont {Broglia}}, \bibinfo {author} {\bibfnamefont {A.}~\bibnamefont {Brunengo}}, \emph {et~al.},\ }\href {https://doi.org/https://doi.org/10.1016/S0168-9002(03)01368-8} {\bibfield  {journal} {\bibinfo  {journal} {Nucl. Instrum. Methods Phys. Res., Sect. A}\ }\textbf {\bibinfo {volume} {506}},\ \bibinfo {pages} {250} (\bibinfo {year} {2003})}\BibitemShut {NoStop}%
\bibitem [{\citenamefont {Kowalski}\ and\ \citenamefont {Enge}(1986)}]{RAYTRACE}%
  \BibitemOpen
  \bibfield  {author} {\bibinfo {author} {\bibfnamefont {S.}~\bibnamefont {Kowalski}}\ and\ \bibinfo {author} {\bibfnamefont {H.~A.}\ \bibnamefont {Enge}},\ }\href {https://github.com/eames/mdmsim/blob/master/MDMTrace/src/RAYTKIN1.F} {\bibinfo {title} {Raytrace code}} (\bibinfo {year} {1986})\BibitemShut {NoStop}%
\bibitem [{\citenamefont {Erell}\ \emph {et~al.}(1986)\citenamefont {Erell}, \citenamefont {Alster}, \citenamefont {Lichtenstadt}, \citenamefont {Moinester}, \citenamefont {Bowman}, \citenamefont {Cooper}, \citenamefont {Irom}, \citenamefont {Matis}, \citenamefont {Piasetzky},\ and\ \citenamefont {Sennhauser}}]{Semi}%
  \BibitemOpen
  \bibfield  {author} {\bibinfo {author} {\bibfnamefont {A.}~\bibnamefont {Erell}}, \bibinfo {author} {\bibfnamefont {J.}~\bibnamefont {Alster}}, \bibinfo {author} {\bibfnamefont {J.}~\bibnamefont {Lichtenstadt}}, \bibinfo {author} {\bibfnamefont {M.~A.}\ \bibnamefont {Moinester}}, \bibinfo {author} {\bibfnamefont {J.~D.}\ \bibnamefont {Bowman}}, \bibinfo {author} {\bibfnamefont {M.~D.}\ \bibnamefont {Cooper}}, \bibinfo {author} {\bibfnamefont {F.}~\bibnamefont {Irom}}, \bibinfo {author} {\bibfnamefont {H.~S.}\ \bibnamefont {Matis}}, \bibinfo {author} {\bibfnamefont {E.}~\bibnamefont {Piasetzky}},\ and\ \bibinfo {author} {\bibfnamefont {U.}~\bibnamefont {Sennhauser}},\ }\href {https://doi.org/10.1103/PhysRevC.34.1822} {\bibfield  {journal} {\bibinfo  {journal} {Phys. Rev. C}\ }\textbf {\bibinfo {volume} {34}},\ \bibinfo {pages} {1822} (\bibinfo {year} {1986})}\BibitemShut {NoStop}%
\bibitem [{\citenamefont {Hyldegaard}\ \emph {et~al.}(2009)\citenamefont {Hyldegaard}, \citenamefont {Forss\'en}, \citenamefont {Diget}, \citenamefont {Alcorta}, \citenamefont {Barker}, \citenamefont {Bastin}, \citenamefont {Borge}, \citenamefont {Boutami}, \citenamefont {Brandenburg}, \citenamefont {B\"uscher}, \citenamefont {Dendooven}, \citenamefont {{Van Duppen}}, \citenamefont {Eronen}, \citenamefont {Fox}, \citenamefont {Fulton} \emph {et~al.}}]{fynbo-2009}%
  \BibitemOpen
  \bibfield  {author} {\bibinfo {author} {\bibfnamefont {S.}~\bibnamefont {Hyldegaard}}, \bibinfo {author} {\bibfnamefont {C.}~\bibnamefont {Forss\'en}}, \bibinfo {author} {\bibfnamefont {C.}~\bibnamefont {Diget}}, \bibinfo {author} {\bibfnamefont {M.}~\bibnamefont {Alcorta}}, \bibinfo {author} {\bibfnamefont {F.}~\bibnamefont {Barker}}, \bibinfo {author} {\bibfnamefont {B.}~\bibnamefont {Bastin}}, \bibinfo {author} {\bibfnamefont {M.}~\bibnamefont {Borge}}, \bibinfo {author} {\bibfnamefont {R.}~\bibnamefont {Boutami}}, \bibinfo {author} {\bibfnamefont {S.}~\bibnamefont {Brandenburg}}, \bibinfo {author} {\bibfnamefont {J.}~\bibnamefont {B\"uscher}}, \bibinfo {author} {\bibfnamefont {P.}~\bibnamefont {Dendooven}}, \bibinfo {author} {\bibfnamefont {P.}~\bibnamefont {{Van Duppen}}}, \bibinfo {author} {\bibfnamefont {T.}~\bibnamefont {Eronen}}, \bibinfo {author} {\bibfnamefont {S.}~\bibnamefont {Fox}}, \bibinfo {author} {\bibfnamefont {B.}~\bibnamefont {Fulton}}, \emph {et~al.},\ }\href {https://doi.org/https://doi.org/10.1016/j.physletb.2009.06.064} {\bibfield  {journal} {\bibinfo  {journal} {Phys. Lett. B}\ }\textbf {\bibinfo {volume} {678}},\ \bibinfo {pages} {459} (\bibinfo {year} {2009})}\BibitemShut {NoStop}%
\bibitem [{\citenamefont {Munch}\ \emph {et~al.}(2016)\citenamefont {Munch}, \citenamefont {Alcorta}, \citenamefont {Fynbo}, \citenamefont {Albers}, \citenamefont {Almaraz-Calderon}, \citenamefont {Avila}, \citenamefont {Ayangeakaa}, \citenamefont {Back}, \citenamefont {Bertone}, \citenamefont {Carnelli}, \citenamefont {Carpenter}, \citenamefont {Chiara}, \citenamefont {Clark}, \citenamefont {DiGiovine}, \citenamefont {Greene} \emph {et~al.}}]{fynbo-2016}%
  \BibitemOpen
  \bibfield  {author} {\bibinfo {author} {\bibfnamefont {M.}~\bibnamefont {Munch}}, \bibinfo {author} {\bibfnamefont {M.}~\bibnamefont {Alcorta}}, \bibinfo {author} {\bibfnamefont {H.~O.~U.}\ \bibnamefont {Fynbo}}, \bibinfo {author} {\bibfnamefont {M.}~\bibnamefont {Albers}}, \bibinfo {author} {\bibfnamefont {S.}~\bibnamefont {Almaraz-Calderon}}, \bibinfo {author} {\bibfnamefont {M.~L.}\ \bibnamefont {Avila}}, \bibinfo {author} {\bibfnamefont {A.~D.}\ \bibnamefont {Ayangeakaa}}, \bibinfo {author} {\bibfnamefont {B.~B.}\ \bibnamefont {Back}}, \bibinfo {author} {\bibfnamefont {P.~F.}\ \bibnamefont {Bertone}}, \bibinfo {author} {\bibfnamefont {P.~F.~F.}\ \bibnamefont {Carnelli}}, \bibinfo {author} {\bibfnamefont {M.~P.}\ \bibnamefont {Carpenter}}, \bibinfo {author} {\bibfnamefont {C.~J.}\ \bibnamefont {Chiara}}, \bibinfo {author} {\bibfnamefont {J.~A.}\ \bibnamefont {Clark}}, \bibinfo {author} {\bibfnamefont {B.}~\bibnamefont {DiGiovine}}, \bibinfo {author} {\bibfnamefont {J.~P.}\ \bibnamefont {Greene}}, \emph {et~al.},\ }\href {https://doi.org/10.1103/PhysRevC.93.065803} {\bibfield  {journal} {\bibinfo  {journal} {Phys. Rev. C}\ }\textbf {\bibinfo {volume} {93}},\ \bibinfo {pages} {065803} (\bibinfo {year} {2016})}\BibitemShut {NoStop}%
\end{thebibliography}%

\end{document}